\documentclass[twocolumn,twocolappendix]{aastex631}

\usepackage{amsmath}
\usepackage{empheq}
\usepackage{graphicx}
\expandafter\let\csname equation*\endcsname\relax
  \expandafter\let\csname endequation*\endcsname\relax 
\usepackage{mathrsfs}
\usepackage{textcomp}
\usepackage{enumitem}   
\usepackage{gensymb}

\usepackage{graphicx}
\usepackage[caption=false]{subfig}
\usepackage{multirow}
\usepackage{longtable}
\usepackage[flushleft]{threeparttable}
\usepackage{booktabs} 
\usepackage{CJK}
\usepackage{verbatim}
\usepackage{amssymb}

\usepackage{hyperref}
\hypersetup{colorlinks, linkcolor={blue}, citecolor={blue}, urlcolor={blue}}

\definecolor{DarkOrange}{RGB}{204, 85, 0}
\definecolor{LincolnGreen}{RGB}{17, 102, 0}

\usepackage{xspace}

\newcommand\swift{\textit{Swift}\xspace}
\newcommand\xmm{\textit{XMM-Newton}\xspace}

\newcommand\W {{W^r_{\ \phi}}}

\newcommand\txmm{3XMM~J2150$-$05\xspace}

\defcitealias{Patra2025}{P25}

\begin{document}
\pagenumbering{arabic}

\title{The Wandering Supermassive Black Hole Powering the off-nuclear TDE AT2024tvd}

\author[0000-0002-5063-0751]{M. Guolo}
\affiliation{Bloomberg Center for Physics and Astronomy, Johns Hopkins University, 3400 N. Charles St., Baltimore, MD 21218, USA}
\author{A. Mummery}
\affiliation{School of Natural Sciences, Institute for Advanced Study, 1 Einstein Drive, Princeton, NJ 08540, USA }

\author[0000-0002-3859-8074]{S. van Velzen}
\affiliation{Leiden Observatory, Leiden University, Postbus 9513, 2300 RA Leiden, NL }

\author[0000-0002-2555-3192]{M. Nicholl}
\affiliation{Astrophysics Research Centre, School of Mathematics and Physics, Queens University Belfast, Belfast BT7 1NN, UK}

\author[0000-0003-3703-5154]{S. Gezari}
\affiliation{Department of Astronomy, University of Maryland, College Park, MD, 20742-2421, USA}

\author[0000-0001-6747-8509]{Y. Yao}
\affiliation{Miller Institute for Basic Research in Science, 206B Stanley Hall, Berkeley, CA 94720, USA}
\affiliation{Department of Astronomy, University of California, Berkeley, CA 94720-3411, USA}

\author[0000-0001-6965-7789]{K. C. Chambers}
\affiliation{Institute for Astronomy, University of Hawaii, 2680 Woodlawn Dr., Honolulu, HI 96822, USA}

\author[0000-0001-5486-2747]{T. de Boer}
\affiliation{Institute for Astronomy, University of Hawaii, 2680 Woodlawn Dr., Honolulu, HI 96822, USA}

\author[0000-0003-1059-9603]{M. E. Huber}
\affiliation{Institute for Astronomy, University of Hawaii, 2680 Woodlawn Dr., Honolulu, HI 96822, USA}

\author[0000-0002-7272-5129]{C.-C. Lin}
\affiliation{Institute for Astronomy, University of Hawaii, 2680 Woodlawn Dr., Honolulu, HI 96822, USA}

\author[0000-0002-9438-3617]{T. B. Lowe}
\affiliation{Institute for Astronomy, University of Hawaii, 2680 Woodlawn Dr., Honolulu, HI 96822, USA}

\author[0000-0002-7965-2815]{E. A. Magnier}
\affiliation{Institute for Astronomy, University of Hawaii, 2680 Woodlawn Dr., Honolulu, HI 96822, USA}

\author[0000-0002-6639-6533]{G. Paek}
\affiliation{Institute for Astronomy, University of Hawaii, 2680 Woodlawn Dr., Honolulu, HI 96822, USA}

\author[0000-0002-1341-0952]{R. Wainscoat}
\affiliation{Institute for Astronomy, University of Hawaii, 2680 Woodlawn Dr., Honolulu, HI 96822, USA}

\begin{abstract}

We present an analysis of the spectral energy distribution (SED) of the off-nuclear tidal disruption event (TDE) AT2024tvd during its late-time plateau phase, combining X-ray spectra and UV/optical photometry. Using a fully relativistic, compact accretion disk model with self-consistent inner-disk Comptonization, we reproduce the observed SED without significant residuals. The inferred black hole mass ${\rm log}_{10}(M_{\bullet}/M_\odot) \approx 6.0 \pm 0.2$, and the inferred disk parameters place AT2024tvd within known TDE-disk scaling relations ($L_{\rm bol}^{\rm disk}/L_{\rm Edd} \propto T_{\rm p}^4 \propto M_{\bullet}^{-1}$, $L_{\rm plat} \propto M_{\bullet}^{2/3}$, $R_{\rm out}/r_{\rm g} \propto M_{\bullet}^{-2/3}$).
Our results show that: (i) there is no \textit{currently detected} star cluster or dwarf galaxy associated with the source, down to a mass limit of $\log_{10}(M_{\rm gal}/M_{\odot}) \leq 7.6$; (ii) the black hole is a wandering supermassive, rather than intermediate-mass, black hole; and (iii) the source represents an extreme case of black hole–to–host mass ratio, with $M_{\bullet}/M_{\rm gal} > 3\%$, consistent with a heavily tidally stripped nucleus. The latter aligns with cosmological simulations predicting that surviving host remnants of most wandering black holes should not retain a detectable stellar overdensity when located at small halo-centric distances. We discuss differences with previous analyses of this source and highlight why our modeling approach provides a more physically consistent solution with more reliable parameter inference.

\end{abstract}


\vspace{1em}

\section{Introduction}
The scattering of a star within the radius of influence of a black hole onto a radial orbit that crosses its tidal radius can produce a luminous tidal disruption event \citep[TDE;][]{rees,Gezari2021}. Such events are among the few observational probes of the existence and properties of otherwise quiescent BHs at distances beyond those of the nearest galaxies, where dynamical modeling of stars and gas is still feasible. We have now entered the era of population-level studies of TDEs, with more than 100 candidates and dozens of extremely well-observed sources whose basic multi-wavelength emission properties have been mapped out \citep[e.g.,][]{van_Velzen_21,Charalampopoulos2022,Yao2023,Guolo2024,cendes2024,Mummery2024,Grotova2025,Mummery_vanVelzen2024}.

Inferring the properties of a black hole, particularly its mass ($M_{\bullet}$), directly from the emission produced by a TDE has proven challenging. Physical models developed to explain the poorly understood early-time optical flare have systematically failed \citep[see][]{Hammerstein2022,Ramsden2022,Guolo2025d} to recover, at a statistically significant level, known correlations between the derived $M_{\bullet}$ values and host galaxy properties, including total stellar mass ($M_{\rm gal}$), bulge mass ($M_{\rm bulge}$), and velocity dispersion ($\sigma_{\star}$) \citep{Gultekin2009,Kormendy2013,Reines2015,Greene2020}.

In contrast, methods based on the late-time UV/optical plateau \citep{Mummery2024} and/or the X-ray emission \citep{Guolo_Mummery2025,Guolo2025d}, both known to be powered by direct emission from the newly formed (expanding) accretion disk \citep{Cannizzo1990,Mummery2020}, have been shown to independently reproduce, with high statistical significance, all three correlations: $M_{\bullet}$–$M_{\rm gal}$, $M_{\bullet}$–$M_{\rm bulge}$, and $M_{\bullet}$–$\sigma_{\star}$ \citep{Mummery2024,Ramsden2025,Guolo2025d}.

\begin{figure*}[t!]
    \centering
    \includegraphics[width=\linewidth]{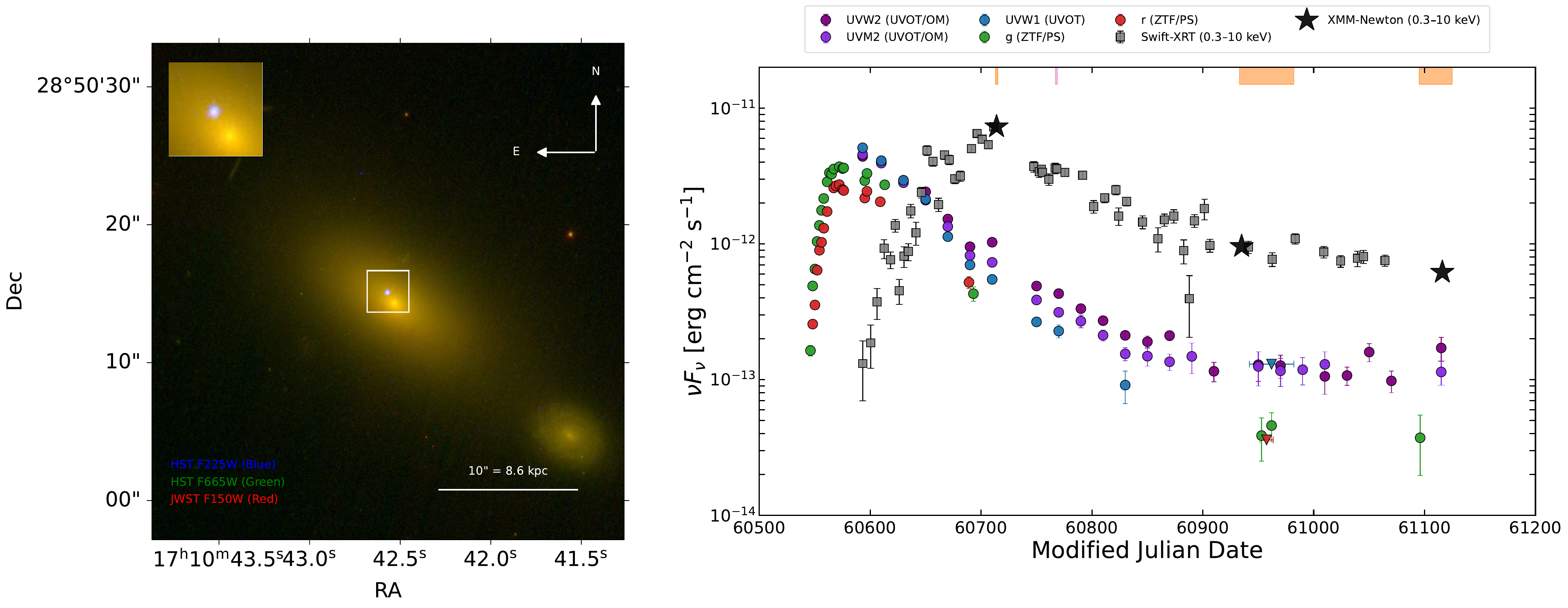}
    \caption{Left: diffraction-limited RGB image of AT2024tvd using HST+JWST imaging. Right: multi-wavelength light curves of AT2024tvd, with the orange marker showing the epochs at which our SED fitting was performed. For plotting purposes, the UVOT light curves were binned to have no more than one point every five days.}
    \label{fig:1}
\end{figure*}

Such methods, which can derive $M_{\bullet}$ values directly from the observed emission itself—without \textit{assuming} any host-scaling relation but instead aiming to probe and populate such correlations—are essential as we begin to unveil a population of TDEs in environments where these scaling relations are not applicable.

The discovery of \txmm \citep{Lin2018,Lin2020} marked the first off-nuclear TDE; the source is associated with an ultra-compact dwarf (UCD) galaxy with $\log_{10}(M_{\rm gal}/M_{\odot}) = 7.3 \pm 0.3$, located in the outskirts of a much more massive early-type galaxy ($M_{\star}\sim10^{11} \ M_{\odot}$). Another such UCD/off-nuclear TDE, EP240222a, with very similar characteristics, e.g., $\log_{10}(M_{\rm gal}/M_{\odot}) \sim 7.0$, was discovered by the \textit{Einstein Probe} mission \citep{Yuan2018} and reported by \cite{Jin2025}. Disk modeling of both sources yields black hole masses in the intermediate-mass black hole (IMBH; $10^4 < M_{\bullet}/M_{\odot} \lesssim 10^5$) regime \citep{Jin2025,Guolo2025d}. 

More recently, the first off-nuclear TDE discovered by an optical survey, AT2024tvd \citep{Yao2025}, showed all the classical characteristics of a TDE but is located 0.8 kpc (projected distance) from the nucleus of another massive ($M_{\star} \sim 10^{11} \ M_{\odot}$) early-type galaxy. Distinct from the previous two sources, pre-TDE images at the position of AT2024tvd reveal no clear stellar counterpart other than the halo of the parent galaxy, and modeling and subtraction of the parent galaxy surface brightness profile show no residuals down to a limit of $\log_{10}(M_{\rm gal}/M_{\odot}) \leq 7.6$ \citep{Yao2025}.

The growing number of such sources appears to be opening a new window into the discovery of a population of off-nuclear IMBHs and/or wandering supermassive black holes (SMBHs), long predicted theoretically to exist \citep{Volonteri2008,Ricarte2018,Inayoshi2020,Ricarte2021a}. These are, however, difficult to detect, as gas accretion onto these systems should be negligibly low in the local universe \citep{Ricarte2021b,Ricarte2021a}. Non-accreting WBHs have been inferred dynamically in a small number of ultra-compact dwarf galaxies \citep[UCDs; e.g.,][]{Seth2014,Taylor2025}, which are common in galaxy clusters \citep{Pfeffer2014,Voggel2019}. 
Therefore, the transient high accretion rate reached following a TDE is possibly the most effective way to unambiguously discover and characterize the accreting counterparts of these WBH systems. 

The goals of this paper are to present follow-up observations of AT2024tvd, capturing the transition of its UV/optical light curve to the plateau phase, as well as novel high signal-to-noise (S/N) ratio \xmm\ X-ray spectra. These comprehensive multi-wavelength data facilitate full spectral energy distribution (SED) fitting to infer AT2024tvd's physical properties, including its black hole mass $M_{\bullet}$. The method/model used here has been shown to yield precise, reliable, and self-consistent results, including the ability to reproduce host-scaling relations with high statistical significance \citep{Guolo2025d}. 

This paper is organized as follows. In \S\ref{sec:data}, we present the data and fitting method; in \S\ref{sec:results}, we describe the results of our SED fitting; in \S\ref{sec:discussion}, we discuss these results, including their relation to previous work on the source; and in \S\ref{sec:conclusio}, we provide our conclusions. A Bayesian statistical framework is used throughout the paper, and the converged posteriors of the inferred parameters are reported as medians, while the uncertainties correspond to the 68\% credible intervals, unless otherwise stated.

\section{Data and Fitting}\label{sec:data}
The data used in this work are primarily public, including light curves from the Zwicky Transient Facility \citep[ZTF;][]{Bellm2019}, observations from the Ultraviolet/Optical Telescope (UVOT; \citealp{Roming2005}) and the X-Ray Telescope (XRT; \citealp{Burrows2005}) on board the \textit{Neil Gehrels Swift Observatory} \citep{Gehrels2004}, and diffraction-limited UV/optical/IR imaging from the \textit{Hubble Space Telescope} (\textit{HST}) and the \textit{James Webb Space Telescope} (\textit{JWST}), which are used to produce the RGB composite shown in Figure~\ref{fig:1}. We also present novel proprietary high signal-to-noise-ratio X-ray spectra from \xmm\ (OBSID 0942540601, 0942540901 and 0942541001; PI: Guolo) obtained at three epochs, as well as deep late-time ground-based photometry from Pan-STARRS \citep{Chambers2016}. Details of the data reduction, including host-light subtraction, are provided in Appendix~\ref{app:data}.

\begin{figure*}
    \centering
    \includegraphics[width=0.9\linewidth]{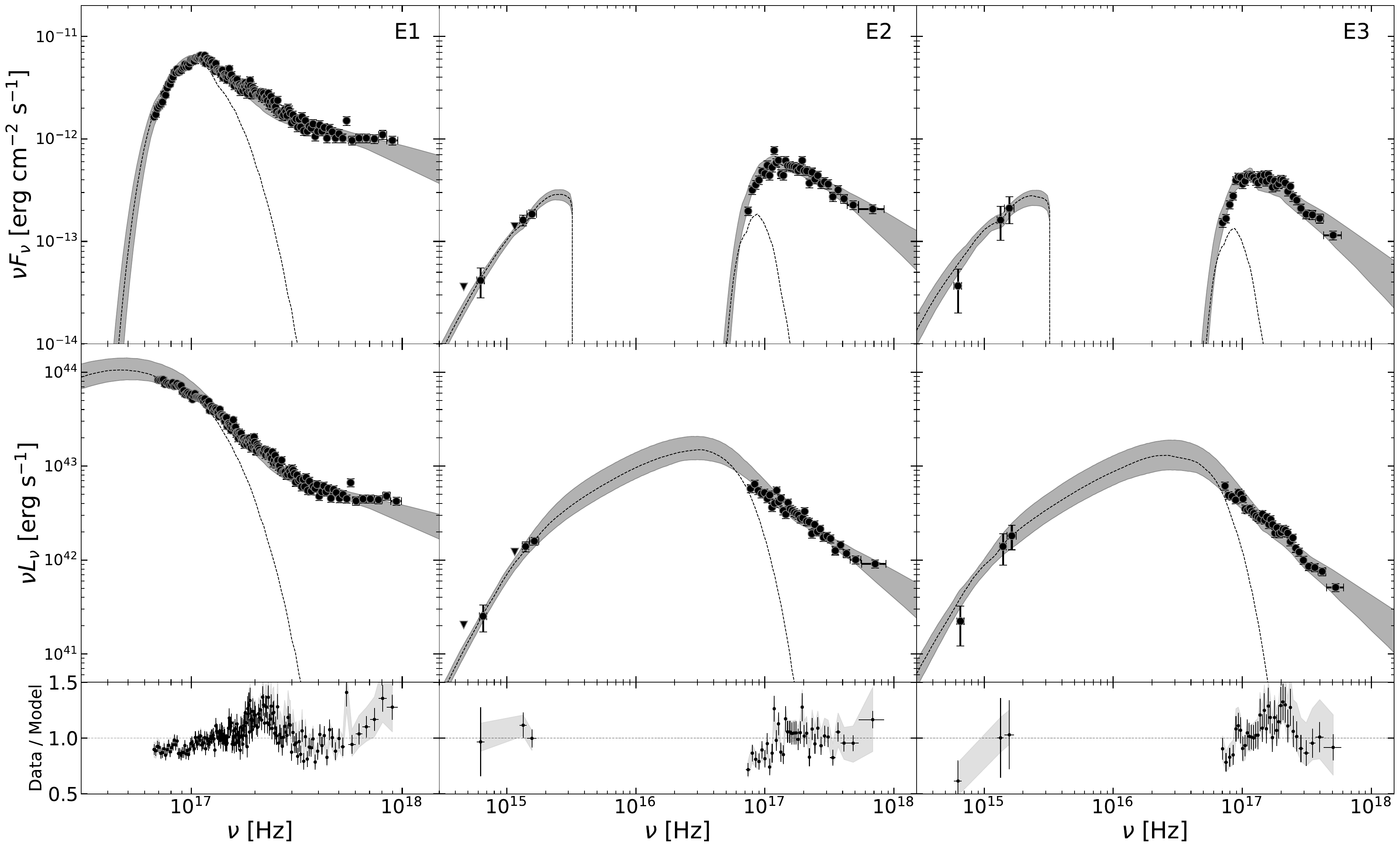}
    \caption{Spectral energy distribution modeling of AT2024tvd. The upper panels show the emission in the observer frame, while the lower panels display the rest-frame emission corrected for all absorption and extinction/attenuation effects. The left column presents the early-time (E1), disk-dominated \xmm\ spectrum, which is well reproduced within an inner accretion disk plus corona framework. At later times (E2 and E3), the full optical/UV–X-ray SED is consistently reproduced using the same disk framework (middle and right columns). The dashed lines indicate the median disk solutions with the coronal component turned off. Bottom panels show plot residuals.}

    \label{fig:2}
\end{figure*}

In the right panel of Fig.~\ref{fig:1}, we show the multi-wavelength light curve of AT2024tvd. Two key properties of the source are now evident from the extended follow-up beyond the data presented in the discovery paper. First, the transient appears to have reached its plateau phase \citep{Mummery2024} around 250 days after discovery ($\Delta t \sim 250$ days). This is seen clearly in the UV light curve, which decayed by a factor of $\sim10$ during the first 100 days after peak (MJD 60600–60700), by an additional factor of $\lesssim10$ over the following 100 days (MJD 60700–60800), and then remained statistically consistent with being constant over the next 200 days (MJD 60900–61100).

Second, the X-ray peak is now clearly identified as having occurred around $\Delta t \sim 150$ days, or approximately 100 days after the optical/UV peak—behavior that is not uncommon among TDEs \citep[e.g.,][]{Gezari2017,Guolo2024}. Our first \xmm\ epoch appears to coincide closely with this X-ray peak. By the time of the second and third \xmm\ epochs, the source had already entered the plateau phase in its UV/optical light curve. At these epochs, ZTF no longer detected the transient, but our deep Pan-STARRS follow-up securely detected the source in the $g$-band and placed a deep upper limit in the $r$-band. The Pan-STARRS observations are quasi-simultaneous with the second and third \xmm\ epochs, taken no more than 15 days apart. During the time of the third \xmm\ epoch, \swift\ had already been turned off in preparation for its expected reboot, so we use \xmm\ Optical Monitor (OM) data to complement the UV follow-up.

To derive the physical properties of AT2024tvd, we perform full SED fitting of the X-ray spectra and UV/optical photometry using the \texttt{kerrSED} model \citep{Guolo_Mummery2025}. This model describes a relativistic, color-corrected, quasi-steady-state accretion disk with a vanishing-stress inner boundary condition, and includes five free parameters: the inner disk radius ($R_{\rm in}$), peak disk temperature ($T_{\rm p}$), outer disk radius ($R_{\rm out}$), black hole spin ($a_\bullet$), and inclination ($i$). Implemented within the \texttt{python} interface to \texttt{XSPEC} \citep{Arnaud_96}, \texttt{kerrSED} enables simultaneous multi-wavelength fitting while performing numerical ray tracing on the fly to capture relativistic effects self-consistently. Relative to standard soft/thermal spectral models, \texttt{kerrSED} provides several key improvements: it incorporates relativistic photon trajectory corrections, treats the disk size as a free parameter --- essential for TDE applications --- and includes a temperature-dependent color correction to account for radiative transfer effects \citep{Shimura1995,Hubeny2001,Davis19}. These features are critical for obtaining physically reliable parameter constraints from continuum fitting.

For a more complete description of the model, including its assumptions, limitations, comparison to other commonly used \texttt{XSPEC} models, and demonstrations of the influence of each parameter on the emergent emission, we refer the reader to \citet{Guolo_Mummery2025}. Applications of \texttt{kerrSED} to a broader TDE sample are presented in \citet{Guolo2025d}. Particularly relevant here is that fitting a sample of 16 TDEs --- including 15 hosted by massive galaxies with $9.5 \lesssim \log_{10}(M_{\star}/M_{\odot}) \lesssim 10.7$ and the UCD hosting \txmm\ --- recovered the $M_{\bullet}$--$M_{\rm host}$ and $M_{\bullet}$--$\sigma_{\star}$ relations with high statistical significance. This result shows that the inferred $M_{\bullet}$ values with this approach are not merely preferred values for model parameters, but instead trace physically meaningful black hole properties. It also enables the properties of AT2024tvd to be directly compared against a uniformly modeled TDE sample.

Following \citet{Guolo2025d}, we use this model to fit the two epochs (E2 and E3) of X-ray spectra plus UV/optical photometry during the plateau phase, as well as another epoch (E1) of X-ray spectra \textit{only}, corresponding to the time when the X-rays were near peak flux. The rationale behind these choices is discussed in detail by \citet{Guolo2025d}, but can be summarized based on several key aspects of TDEs. First, the UV/optical plateau is produced by direct emission from the outer regions of a compact, expanding disk \citep{Mummery2024}. Second, the same disk powers the X-ray emission at all times \citep[as shown independently through individual source fitting and luminosity function analysis; see][]{Mummery2020, Mummery_vanVelzen2024,Guolo2025d}. Third, the early-time optical emission (prior to the plateau) is \emph{not} produced by direct disk emission, and cannot be modeled with such models. These facts imply that during the plateau phase, the entire SED is generated by direct disk emission and can therefore be reliably modeled with such models, but this is not true prior to the plateau. Nevertheless, the peak X-ray epoch, being sourced by the same disk, still aids in constraining black hole properties, particularly helping to break degeneracies between dynamical and intrinsic parameters \citep[see discussion in][]{Guolo2025d}.

For the X-rays, we use three \textit{XMM-Newton} spectra that satisfy the above conditions. For the second (third) epoch, we include both UVOT (OM) and Pan-STARRS data (including both detections and upper limits). These three epochs are marked in orange in Figure~\ref{fig:1}. The fitting is carried out using the Bayesian X-ray Analysis software (BXA) version 4.0.7 \citep{Buchner2014}, which connects the nested sampling algorithm \texttt{UltraNest} \citep{Buchner2019} with the \texttt{PyXspec} fitting environment. The three epochs are loaded and fitted simultaneously under the assumption of Gaussian statistics. The Gaussian statistics assumption is justified because the X-ray spectra were binned to have a minimum background-subtracted ${\rm S/N}$ of 10, in each bin, which, combined with the $\geq 1000$ counts (see Appendix \ref{app:data}) in each spectrum, ensures that each bin is comfortably in the Gaussian regime.

In addition to the disk model, we account for both Galactic --- using \texttt{phabs$\times$redden} and assuming known values for column density and color excess \citep{heasarcnh,Schlafly2011} --- and intrinsic (using \texttt{phabs$\times$reddenSF}) absorption and reddening due to gas and dust along the line of sight. The source emission is redshifted to the measured redshift $z$ using \texttt{zashift}. In \texttt{XSPEC} notation, the model is 
\texttt{phabs$\times$redden$\times$zashift(phabs$\times$reddenSF$\times$(SimPL$\otimes$} \texttt{kerrSED))}, 
where \texttt{SimPL} \citep{Steiner2009} is used to self-consistently account for strong Comptonization (i.e., the hard X-ray tail), while conserving the number of photons—an important feature of AT2024tvd's X-ray spectra \citep{Yao2025}.

We adopt the same linear or log-linear priors and parameter limits as described in \citet{Guolo2025d} and shown in their Table 3. In the fitting process, the five free parameters of \texttt{kerrSED}---$R_{\rm in}$, $T_{\rm p}$, $R_{\rm out}$, $a_\bullet$, and $i$---are treated as follows: $T_{\rm p}$ and $R_{\rm out}$, being dynamical, are allowed to vary between epochs, while the remaining parameters are tied between the two epochs. However, since the first epoch includes X-ray data only (as the early-time optical flare is not powered by the disk), $R_{\rm out}(E1)$ is completely unconstrainable, and is simply assumed to be $R_{\rm out}(E1) \leq R_{\rm out}(E_N)$ for $N>1$, a physical prior resulting from angular momentum conservation \citep{Cannizzo1990,Mummery2020}. The corona/\texttt{SimPL} parameters ($f_{sc}$ and $\Gamma$) are treated as independent and free parameters. The posteriors on black hole mass and disk bolometric luminosity ($L_{\rm Bol}^{\rm disk}$) are then extracted from the posteriors of the free parameters; see Equations 3 and 4 in \citet{Guolo2025d}.

\section{Results}\label{sec:results}

The resulting SED fitting is shown in Fig.~\ref{fig:2}. The left panel presents the first \xmm\ epoch, obtained near the X-ray peak, where the spectrum is well described by a hot inner disk with a modest contribution from a Comptonizing corona. The midle and right panels show the second and third epochs, corresponding to when the UV/optical emission has entered the plateau phase, is fully accounted for by emission from a compact disk whose inner regions are now dominated by stronger Comptonization.  

Crucially,  for the discussion in \S\ref{sec:discussion}, as shown the bottom panels the combined X-ray spectra and UV/optical SED can be modeled with no significant residuals (particularly in the UV/optical) in the disk-dominated plateau-phase epochs. This indicates that the disk (plus corona) emission alone can explain all of the observed, pre-transient--subtracted flux during the second and third epochs. The median values and $1\sigma$ uncertainties of the inferred parameters are summarized in Table~\ref{tab:best_fit}, while the full posterior distributions are shown in Appendix~\ref{app:post}.

\begin{deluxetable*}{c|ccccc|ccc}
\tablecaption{Summary of Inferred Parameters \label{tab:best_fit}}
\tablehead{
\colhead{$\Delta t$ } &  
\colhead{$\log_{10}(N_{\rm H})$} & \colhead{$E(B-V)^{(a)}$} & \colhead{$\log_{10}(R_{\rm in})$} &
\colhead{$\log_{10}(T_{\rm p})$} & \colhead{$\log_{10}(R_{\rm out})$} & \colhead{$\log_{10}(M_{\rm \bullet})$} &
\colhead{$\log_{10}(L_{\rm Bol}^{\rm disk})$} & \colhead{$\log_{10}(L_{\rm Bol}^{\rm disk}/L_{\rm Edd})$} \\
 \colhead{(days)} & \colhead{(cm$^{-2}$)} & \colhead{} & \colhead{(km)} &
\colhead{(K)} & \colhead{($r_g$)} & \colhead{($M_{\odot}$)} & \colhead{(erg s$^{-1}$)} & \colhead{}
}
\startdata
168 & \multirow{3}{*}{$20.57^{+0.20}_{-0.24}$} & \multirow{3}{*}{$0.04^{+0.03}_{-0.02}$} & \multirow{3}{*}{$6.93^{+0.08}_{-0.07}$}  & $5.70^{+0.04}_{-0.04}$ & -- &  \multirow{3}{*}{$5.98^{+0.18}_{-0.15}$} & $44.27^{+0.11}_{-0.12}$ & $0.18^{+0.08}_{-0.09}$ \\
389--436   &  &  &  & $5.50^{+0.05}_{-0.05}$ & $2.81^{+0.15}_{-0.15}$  &  & $43.46^{+0.15}_{-0.11}$ & $-0.62^{+0.14}_{-0.12}$ \\
547--577   &  &  &  & $5.48^{+0.05}_{-0.04}$ & $2.96^{+0.26}_{-0.19}$  &  & $43.38^{+0.16}_{-0.15}$ & $-0.71^{+0.16}_{-0.14}$  \\
\hline
\enddata
\tablecomments{$\Delta t$ refers to the time since the first detection of the source. (a) tied to $N_{\rm H}$ assuming a Galactic-like gas-to-dust ratio.} 
\end{deluxetable*}

\begin{figure*}
    \centering
    \includegraphics[width=0.9\linewidth]{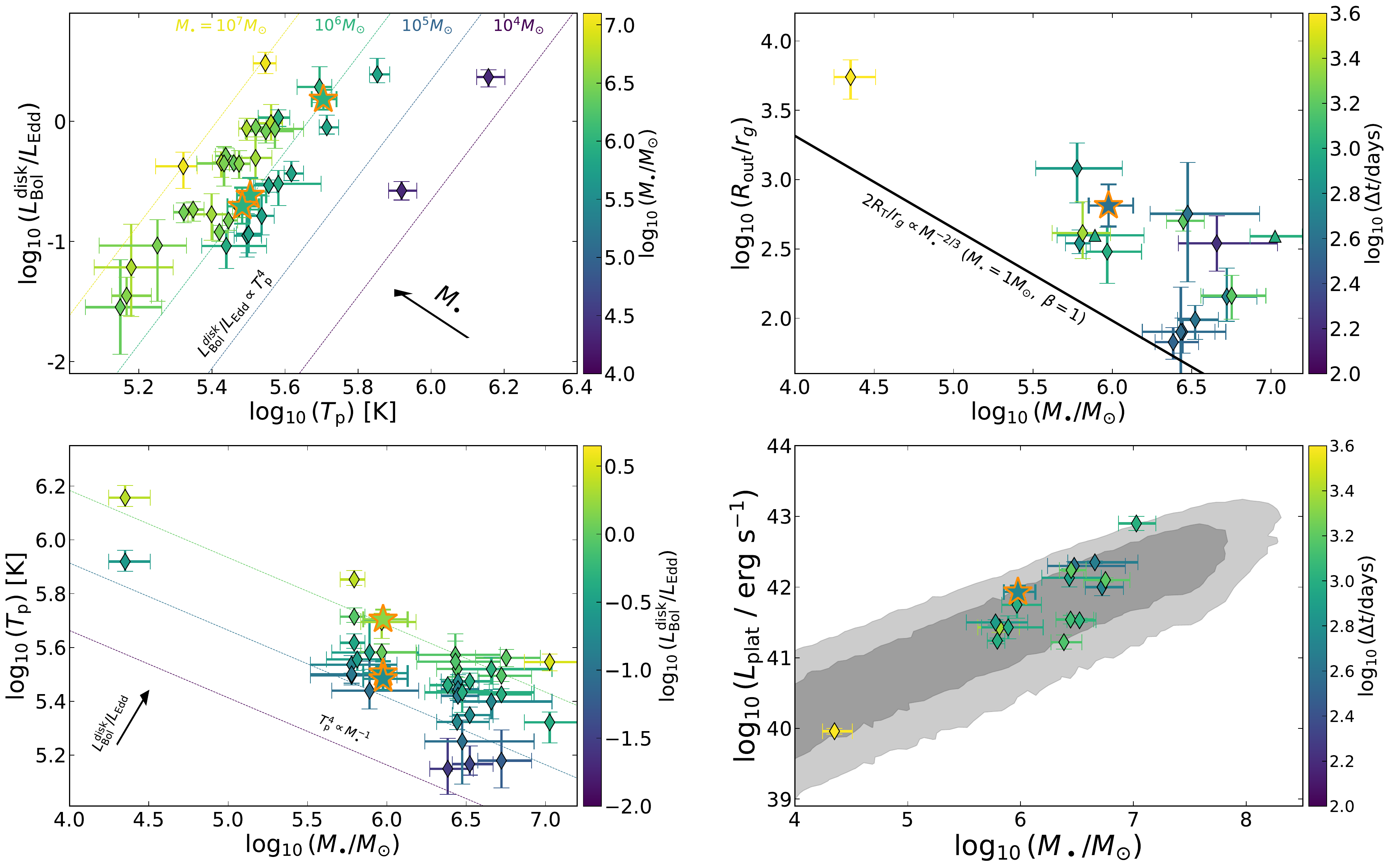}
    \caption{The location of AT2024tvd (points displayed by stars with orange outlines) in the phase space spanned by TDE accretion disks (all other points; presented in \citealt{Guolo2025d} and derived with the same method and model). AT2024tvd has an accretion flow that is entirely consistent with the nuclear TDE population, i.e., powered by an SMBH. 
    Left panels show standard accretion correlations between Eddington ratio, peak disk temperature, and black hole mass. In these panels, the lines of constant $M_{\bullet}$ ($= 10^4$, $10^5$, $10^6$ and $10^7 M_{\odot}$) and constant $L_{\rm Bol}^{\rm disk}/L_{\rm Edd}$ ($=0.01$, $0.1$, $1.0$) are computed for the case of $a_{\bullet}=0$ and $R_{\rm out}=100 r_{\rm g}$. The right panels show correlations that are specific to TDE disks: between the UV luminosity of the plateau and the black hole mass \citep{Mummery2024}, and between the relative size of the disk and the black hole mass. Every scaling relationship can be understood with respect to time-dependent disk theory (see text).  }
    \label{fig:3}
\end{figure*}

From the fitting, we obtain, importantly, $\log_{10}(R_{\rm in}/{\rm km}) \approx 6.9 \pm 0.1$, $\left| a_{\bullet} \right| \lesssim 0.8$ (90\% credible interval), and $i \lesssim 50 \deg$ (90\% credible interval). From these, we infer a black hole mass of $\log_{10}(M_{\bullet}/M_{\odot}) = 5.98^{+0.18}_{-0.14}$. This result is illustrated in Fig.~\ref{fig:3}, where AT2024tvd is placed within the accretion disk parameter space of other TDE systems (alongside 16 other events modeled with the same relativistic disk framework, \texttt{kerrSED}).

In Fig.~\ref{fig:3}, the left panels show standard accretion disk correlations, $L_{\rm bol}^{\rm disk}/L_{\rm Edd} \propto T_{\rm p}^4 \propto M_{\bullet}^{-1}$, while the right panels display TDE-specific correlations, $L_{\rm plat} \propto M_{\bullet}^{2/3}$ and $R_{\rm out}/r_{\rm g} \propto M_{\bullet}^{-2/3}$. In the upper-right panel, the earliest measurements of $R_{\rm out}/r_{\rm g}$ for each source are shown, color-coded by time since discovery. The solid black line marks the circularization radius, $R_{\rm circ} \propto M_{\bullet}^{-2/3}$, normalized for the disruption of a solar-mass main-sequence star with an impact parameter $\beta = 1$. The lower-right panel presents the plateau luminosity, $L_{\rm plat}$, computed from the modeled $\nu L_{\nu}$ SED at $\nu = 10^{15}~{\rm Hz}$ and evaluated at the closest epoch to $\Delta t = 1000~{\rm days}$, overlaid on the $10^6$ time-dependent disk solutions from \citet{Mummery2024} at the same epoch and frequency (i.e., their $L_{\rm plat}-M_{\bullet}$ scaling relation; all of these theoretical points are evaluated at 1000 days).

All correlations in Fig.~\ref{fig:3} follow naturally from classical time-dependent disk theory \citep{Lynden-Bell1974,Cannizzo1990,Balbus2017,Mummery2020}. In particular, the relations shown in the right panels emerge when this framework is applied to disks forming near $R_{\rm circ}$ that subsequently expand to conserve angular momentum \citep[see, e.g.,][]{Mummery2024,Mummery2024fitted,Guolo2025b}. In this parameter space, every TDE with $\log_{10}(M_{\bullet}/M_{\odot}) > 5$ (except AT2024tvd) is a nuclear or ``normal'' TDE, while the outlier at $\log_{10}(M_{\bullet}/M_{\odot}) \approx 4.4$ corresponds to the off-nuclear IMBH TDE \txmm, associated with a detected ultra-compact dwarf galaxy (UCD, $M_{\rm gal} \sim 10^{7} M_{\odot}$). 

From its spectral and multi-wavelength properties, AT2024tvd is therefore indistinguishable from classical, nuclear TDEs powered by SMBHs, and unlike those of the only other off-nuclear/IMBH (\txmm) in the \citep{Guolo2025d} sample, confirming that its emission originates from a supermassive, rather than an intermediate-mass, black hole.

\section{Discussion}\label{sec:discussion}

\subsection{Comparison to previous studies}

It is important to compare the results obtained in the previous section with those reported in the two previous analyses of AT2024tvd. The discovery paper by \citet{Yao2025} presented data covering only the early decline of the optical flare and was limited to observations obtained prior to the X-ray peak. In that study, the authors estimated the black hole mass using two different methods.

First, the authors fitted the optical/UV light curve using \texttt{MOSFIT} \citep{Mockler2019}. While the resulting black hole mass estimate is statistically consistent with the value derived in this work, this agreement is not surprising. \texttt{MOSFIT} is known to systematically recover black hole masses clustered around $\log_{10}(M_{\bullet}/M_{\odot}) \sim 6.5 \pm 0.5$ (1$\sigma$), largely independent of host galaxy properties \citep[e.g., left panel Fig. 12 of][]{Guolo2025d}, such that it is unable to recover any of the known black hole--host galaxy scaling relations \citep{Hammerstein2023a,Ramsden2022,Guolo2025d}. This is likely because its assumed luminosity scaling (with the fallback rate) is in strong conflict with the TDE population data \citep{Mummery2025Calorimetry}, implying that the derived masses are not reliable.

Second, the authors employed the \emph{empirical} correlation between the peak optical luminosity of nuclear TDEs and black hole mass derived by \citet{Mummery2024}, obtaining $\log_{10}(M_{\bullet}/M_{\odot}) = 6.9 \pm 0.5$. While this relation is, by construction, consistent with the data on a population level, and provides a useful first-order estimate \citep{Mummery26TDEFLARE}, it suffers from substantial intrinsic scatter, and is, after all, a scaling relation rather than a modeling of the source emission itself. Moreover, there is currently no physical model capable of explaining this correlation in its entirety \citep[see the detailed discussion in][]{Mummery2025Calorimetry}. 

More recently, \citet[][hereafter \citetalias{Patra2025}]{Patra2025} attempted to infer several properties of AT2024tvd, including $M_{\bullet}$, from its multi-wavelength emission using an X-ray–to–IR analysis. Although superficially similar to our approach, their method differs in several key respects, naturally leading to discrepancies between their inferred parameters and those derived here.
\citetalias{Patra2025} modeled the SED of AT2024tvd at $\Delta t \sim 200$ days (pink tick in Fig.~\ref{fig:1}) using a JWST near-IR spectrum, two \textit{Swift}/UVOT UV bands, ZTF $g$- and $r$-band photometry, and the \emph{integrated} 0.3--10~keV luminosity from \textit{Swift}/XRT represented as a single point at $\sim0.6$~keV. The optical photometry was obtained from difference imaging, while the UV fluxes were host-subtracted using a stellar model derived from pre-transient data.

The SED was fit with a steady-state, non-GR-corrected accretion disk model \citep{Shakura1973} plus a cold blackbody component intended to represent dust-reprocessed emission. Because the model could not reproduce the UV emission, the residuals were attributed to an additional stellar component interpreted as a nuclear star cluster. However, this component is inferred solely from UV residuals and is unsupported by any direct JWST or infrared detection.

 \begin{figure}
    \centering
    \includegraphics[width=1\linewidth]{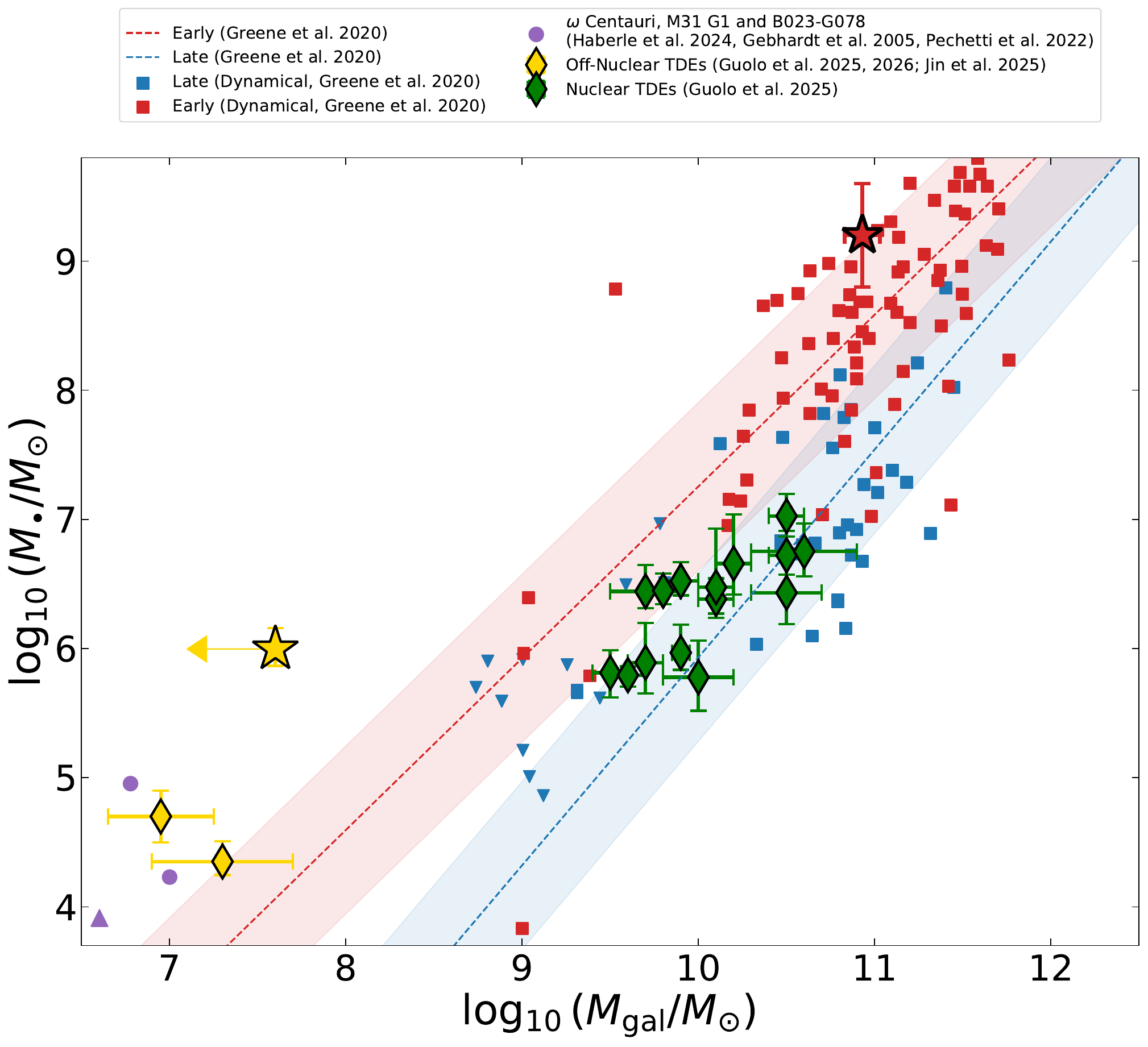}
   \caption{Black hole mass of AT2024tvd (yellow star) in the context of the host–black hole mass relation. Red and blue squares show dynamical black hole mass measurements in nearby galaxies \citep[from][]{Kormendy2013,Greene2020}, with contours indicating the corresponding correlations derived by \citet{Greene2020}. Purple points denote dynamically inferred black holes in globular clusters \citep[from][]{Haberle2024,Gebhardt2005,Pechetti2022}. Nuclear TDEs are shown as green diamonds \citep[from][]{Guolo2025d}, while off-nuclear IMBH TDEs in ultra-compact dwarfs are shown as yellow diamonds (\txmm\ and EP20240222a; \citealt{Lin2018,Guolo2025d,Jin2025}). AT2024tvd is a strong outlier, with only an upper limit on its host stellar mass, consistent with a wandering SMBH embedded in a tidally stripped nucleus. The red star marks the central SMBH of AT2024tvd’s parent galaxy \citep[detected in radio by][]{Yao2025}, with its mass estimated using the $M_{\bullet}$--$\sigma_{\star}$ relation.}
    \label{fig:4}
\end{figure}

This interpretation is problematic because the fit was performed before the source reached the plateau phase. Early-time UV/optical emission in TDEs is not expected to arise from direct disk radiation \citep[e.g.,][]{Gezari2009,vanVelzen2011,Roth2020,Mummery2025Calorimetry}; fitting it with a disk model therefore inevitably produces UV residuals (e.g., Figure 15 of \citealt{Guolo2024}). Only during the plateau phase can the full SED be consistently modeled with a direct disk emission model \citep{Mummery2020,Guolo2025d}. The shallow infrared slope in the JWST data is more likely due to non-thermal free--free emission, as seen in other early-time TDEs \citep{Reynolds2025}. Interpreting this emission as part of a single disk naturally inflates the inferred outer disk radius, explaining the unphysically large value reported by \citetalias{Patra2025}.

Moreover, because the optical/UV data were already host-subtracted, any genuine stellar emission at the TDE location had necessarily been removed. A stellar component therefore cannot be recovered from these data, simply because host light cannot be detected from a host-subtracted light curve or SED. We conclude that the claimed nuclear star cluster is an artifact of fitting an inappropriate model to early-time TDE emission. The upper limits on stellar light at the TDE location derived by \citet{Yao2025} therefore remain valid.

The black hole mass estimate of \citetalias{Patra2025} is also affected by shortcomings in the SED modeling. First, the X-ray emission is reduced to a single luminosity point despite the observed 0.3--10~keV spectrum lying on the exponentially falling Wien tail. In this regime, integrated luminosity alone cannot independently constrain both $\dot{M}_{\bullet}$ and $M_{\bullet}$ without spectral-shape information\footnote{Unless multiple epochs are fit simultaneously, allowing for the degeneracy between dynamical and intrinsic parameters to be broken, as in time-dependent approaches such as \texttt{FitTeD} \citep{Mummery2024fitted}.}. Consequently, the inferred parameters depend sensitively on the arbitrary placement of the X-ray point in the SED, effectively discarding most of the information contained in the X-ray spectra.

Second, the X-ray spectra of AT2024tvd are known not to be purely thermal/disk emission \citep{Yao2025}. Non-thermal Comptonized emission contributes significantly at early times and dominates later on (Fig.~\ref{fig:2}). Reliable mass estimates therefore require models that conserve photon number and account for coronal up-scattering. The model of \citetalias{Patra2025} also neglects color-correction effects required for meaningful disk-based measurements \citep{Shimura1995,Hubeny2001,Davis2006,Davis19,Mummery2021_Xradii}. Ignoring both effects underestimates the intrinsic X-ray flux and thus the normalization parameter, leading to the low mass estimate of $M_{\bullet} \approx 3\times10^5\,M_{\odot}$. Combined with attributing early-time non-disk emission to the disk, this also explains the much larger inferred Eddington ratio. By contrast, we find that AT2024tvd remains close to $L_{\rm Bol}/L_{\rm Edd}\sim1$ even at peak X-ray luminosity, consistent with many TDEs \citep{Mummery2023,Guolo2025d}.
We therefore conclude that the parameters inferred by \citetalias{Patra2025}---including $M_{\bullet}$, $\dot{M}_{\bullet}$, $R_{\rm out}$, and the claimed nuclear star cluster---do not provide reliable measurements of the physical properties of AT2024tvd.

\subsection{Implications for the Origin of AT2024tvd and for Black Hole Demographics}

The inferred mass of the black hole powering AT2024tvd,
$\log_{10}(M_{\bullet}/M_{\odot}) \approx 6.0 \pm 0.2$, provides strong constraints on both the origin of the system and its implications for the broader population of black holes in the local Universe. This measurement places the source firmly within the supermassive regime, distinguishing it from the two previously known off-nuclear TDEs, \txmm\ and EP240222a, which were consistent with IMBHs. AT2024tvd therefore demonstrates that not all off-nuclear TDEs arise from IMBHs still embedded in dwarf satellites in the outskirts of massive galaxies; instead, some originate from displaced, or wandering, SMBHs that have already sunk deep into the potential wells of their halos.

Although we discussed in the previous section how the properties inferred for both the black hole and its stellar counterpart by \citetalias{Patra2025} are inconsistent with our findings, we nevertheless agree with their conclusion that the origin of the black hole is most consistent with a minor merger, rather than with gravitational recoil or dynamical ejection from the parent galactic nucleus. This conclusion follows from a simple statistical argument rooted in hierarchical galaxy formation, in which the number of minor mergers is far greater than the number of major mergers \citep[e.g.,][]{Governato1994,VolonteriBerna2005}. Moreover, the rate of minor mergers is expected to scale with halo mass, which may naturally explain why all three known off-nuclear TDEs have been detected in massive galaxies \citep{Bellovary2010,Ricarte2021a,Ricarte2021b}.
This interpretation is further supported by the long dynamical friction timescales associated with mergers, which are often comparable to or longer than a Hubble time \citep{DosopoulouAntonini2017,DvorkinBarausse2017}. As a result, massive black holes accreted onto massive halos are expected to stall on wide orbits; in the local Universe, most therefore remain offset from the galactic nucleus, with only a small fraction forming close SMBH pairs \citep{Tremmel2015,Tremmel2018binary}. 

The extremely high stellar mass of the parent galaxy of AT2024tvd, its large nuclear velocity dispersion ($\sigma_{\star} \sim 300~\mathrm{km~s^{-1}}$, \citetalias{Patra2025}), and the detection of radio emission from its nucleus \citep{Yao2025} also exclude the possibility that the black hole powering AT2024tvd is the central SMBH of its halo. Using the $M_{\bullet}-\sigma_{\star}$ scaling relations of \citet{Kormendy2013}, the central black hole is instead expected to have a mass of $\log_{10}(M_{\bullet}/M_{\odot}) \approx 9.2 \pm 0.4$, well above the Hills mass (the black hole mass scale at which stars are swallowed whole rather than disrupted; \citealt{Hills_75}).

The location of AT2024tvd—approximately $0.8$~kpc (projected) from the nucleus of a massive early-type galaxy with $M_{\star} \sim 10^{11}~M_{\odot}$—together with the absence of any detected stellar overdensity down to $\log_{10}(M_{\rm gal}/M_{\odot}) \leq 7.6$, strongly suggests that the black hole resides in a stripped remnant nucleus. This is clearly illustrated in Fig.~\ref{fig:4}, which shows the $M_{\bullet}$–$M_{\rm gal}$ plane comparing the dynamical black hole mass compilations of \citet{Kormendy2013} and \citet{Greene2020} with masses inferred from nuclear TDEs \citep[green symbols;][]{Guolo2025d} and from the three off-nuclear TDEs (yellow symbols): EP20240222a, \txmm, and AT2024tvd, with masses taken from \citet{Jin2025}, \citet{Guolo2025d}, and this work, respectively. The black hole masses for all TDEs are derived using the same method and model, except for EP20240222a, which employs a slightly different approach but yields results broadly consistent with \texttt{kerrSED} (see discussion in \citealt{Guolo2025d}). In this figure, the stripped-nucleus nature of AT2024tvd is evident, as the source stands out with an unusually high black hole-to-stellar-mass ratio, $M_{\bullet}/M_{\rm gal} > 0.03$.

This interpretation is fully consistent with predictions from cosmological simulations such as \texttt{ROMULUS} \citep{Tremmel2018wanderings,Ricarte2021a}, which show that hierarchical galaxy assembly naturally produces a population of wandering MBHs that can be broadly—though not exclusively—divided into two classes: black holes that remain far from the halo center and retain some (if not most) of their stellar counterparts, residing in dwarf satellites (as in EP20240222a and \txmm) or even massive globular clusters, and a second population that has migrated deeper into the halo potential but has been stripped of most of its associated (resolved) stellar population and shows no detectable stellar overdensity. This of course does not mean that they have no associated stellar population, e.g., a dense nuclear cluster, which can more easily avoid tidal stripping \citep{Van_Wassenhove2014,Tremmel2018binary} but may be unresolved/undetectable at the available resolution, while still being massive/dense enough to produce TDEs at a detectable rate.

AT2024tvd appears to be the first observed example of this latter population. In a companion paper \citep{Guolo2026}, we show that the properties of off-nuclear TDEs, their host galaxies, and their spatial distribution closely match the predictions of these simulations, even with the currently small sample size.

This discovery has broader implications for black hole demographics. The ability to measure $M_{\bullet}$ directly from the multi-wavelength emission of a TDE—using a physical model that can independently reproduce host scaling relations, but without assuming them—points toward off-nuclear TDEs being likely the only independent way to place constraints on the mass distribution of wandering black holes. More generally, AT2024tvd empirically links the transient accretion population to the wandering black hole population predicted by cosmological simulations, bridging the gap between theoretical models of hierarchical growth and observable time-domain phenomena.

\section{Conclusion}\label{sec:conclusio}

We have presented a detailed analysis of the full spectral energy distribution (SED) of the off-nuclear TDE AT2024tvd during its late-time plateau phase. Using a fully relativistic, compact accretion-disk model with self-consistent inner-disk Comptonization, we reproduce the observed emission without significant residuals. The inferred black hole mass, $\log_{10}(M_{\bullet}/M_{\odot}) \approx 6.0 \pm 0.2$, is robust and places AT2024tvd firmly in the SMBH regime, while remaining fully consistent with established disk-based scaling relations observed in TDEs ($L_{\rm bol}^{\rm disk}/L_{\rm Edd} \propto T_{\rm p}^4 \propto M_{\bullet}^{-1}$, $L_{\rm plat} \propto M_{\bullet}^{2/3}$, $R_{\rm out}/r_{\rm g} \propto M_{\bullet}^{-2/3}$). We also revisited the properties inferred by previous studies and showed why the values derived here are more reliable and better grounded, both observationally and theoretically.

Our results demonstrate that the black hole powering AT2024tvd is fundamentally distinct from those associated with the two previously known off-nuclear TDEs, 3XMM J0251-05 and EP240222a, which were powered by IMBHs ($M_{\bullet} \lesssim 10^5\,M_{\odot}$) and hosted in detected satellite ultra-compact dwarfs. The absence of any detectable stellar overdensity at the transient location, combined with the extreme black hole--to--host mass ratio ($M_{\bullet}/M_{\rm gal} > 3\%$), strongly supports an origin as a highly tidally stripped remnant nucleus hosting a wandering SMBH. This interpretation is consistent with expectations from hierarchical galaxy formation and with cosmological simulations \citep[e.g., \texttt{ROMULUS}][]{Ricarte2021a}, which predict that most wandering black holes at small halo-centric distances should lack detectable surviving stellar counterparts.

More broadly, AT2024tvd demonstrates that off-nuclear TDEs can provide a direct and physically grounded probe of the wandering SMBH population in the local Universe. While upcoming wide-field optical surveys such as the Rubin Observatory Legacy Survey of Space and Time (LSST) are expected to discover many more off-nuclear TDEs, X-ray observations remain essential for achieving reliable and precise black hole mass measurements, as they directly trace emission from the innermost regions of the accretion disk. By enabling black hole mass estimates derived entirely from the transient emission and independent of host-galaxy scaling relations, off-nuclear TDEs offer a unique observational pathway to uncover and characterize displaced black holes that would otherwise remain undetectable.

\vspace{1cm}
\textit{Acknowledgments} -- 
MG is grateful to the Institute for Advanced Study for its hospitality, where part of this work was carried out, and thanks Zachary Lane for help with the color image in Figure 1. MG acknowledges support from NASA through XMM-Newton grant 80NSSC24K1885. A.M. acknowledges support from the Ambrose Monell Foundation, the W.M. Keck Foundation and the John N. Bahcall Fellowship Fund at the Institute for Advanced Study. 
MN is supported by the European Research Council (ERC) under the European Union’s Horizon 2020 research and innovation programme (grant agreement No.~948381). Pan-STARRS is a project of the Institute for Astronomy of the University of Hawaii, and is supported by the NASA SSO Near Earth Observation Program under grants 80NSSC18K0971, NNX14AM74G, NNX12AR65G, NNX13AQ47G, NNX08AR22G, 80NSSC21K1572  and by the State of Hawaii.
Some of the data presented in this article were obtained from the Mikulski Archive for Space Telescopes (MAST) at the Space Telescope Science Institute. The specific observations analyzed can be accessed via \dataset[DOI: 10.17909/hvh9-kr91]{https://doi.org/10.17909/hvh9-kr91}"
\appendix

\section{Data Reduction}\label{app:data}

We performed point-spread-function (PSF) 
photometry on all publicly available ZTF data using the ZTF forced-photometry service \citep{Masci2019,Masci2023} in the $g$ and $r$ bands.

\begin{figure}
    \centering
    \includegraphics[width=\linewidth]{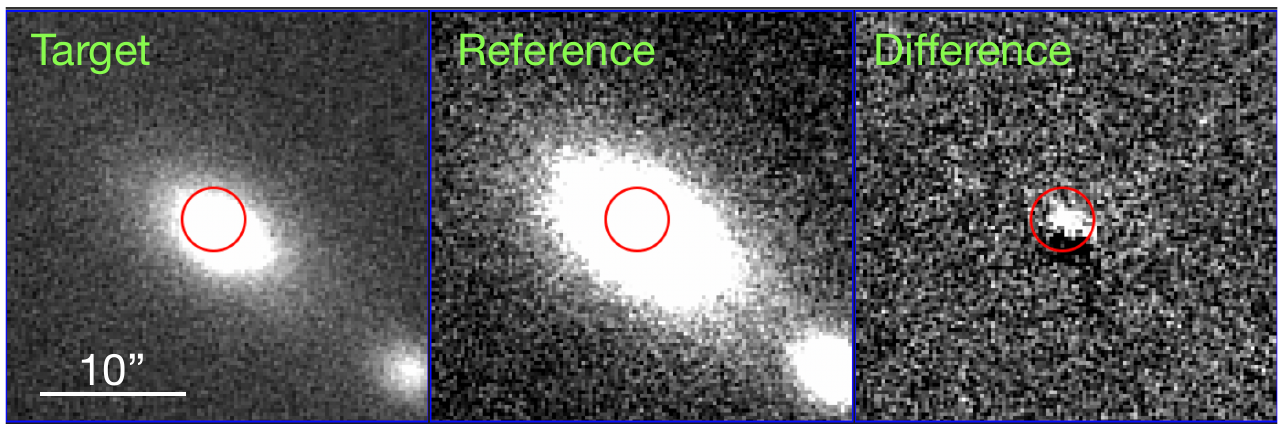}
    \caption{Late-time Pan-STARRS detection of AT2024tvd in $g$-band.}
    \label{fig:ps}
\end{figure}

We observed AT2024tvd using the Panoramic Survey Telescope and Rapid Response System \citep[Pan-STARRS;][]{Chambers2016} on 2025-10-05 and 2025-10-14, with $6\times200$\,s exposures in the Pan-STARRS $g_{\rm P1}$ and $r_{\rm P1}$ filters on each night. All observations were processed and photometrically calibrated with the PS image processing pipeline \citep{Magnier2020,Magnier2020a,Waters2020}, including subtraction of a deep reference image of the same field. The subtraction was complicated by the proximity of AT2024tvd to the bright galaxy nucleus (offset comparable to the full-width at half-maximum of the point-spread function). We downloaded and manually vetted all images, confirming a clear detection of flux in the $g$-band difference images at the position of AT2024tvd, which is shown in Fig.~\ref{fig:ps}. We calculate the flux in each epoch as the mean flux from the reliable $g$-band subtractions. Negative residuals from the galaxy core prevented a clear detection in the $r$-band, and we report only an upper limit in this filter.

The \swift/UVOT photometry was analyzed using the \texttt{uvotsource} package with a 10\arcsec\xspace aperture, similarly with OM data obtained in the third \xmm observation. Host-galaxy flux was subtracted using estimates from population-synthesis modeling of archival pre-event photometry \citep{Yao2025}.

The 0.3--10\,keV \swift/XRT light curves were generated using the UK Swift Data Centre online XRT data products tool \citep{Evans2007}, based on HEASOFT v6.22 \citep[][]{Arnaud1996}, and binned by observation ID. For Fig.~\ref{fig:1}, count rates were converted to fluxes using the nearest best-fit X-ray spectral model from the three analyzed epochs (see main text).

Our primary data set consists of \xmm observations obtained mainly through AO-23 program 94254 (P.I.: Guolo), targeting deep X-ray follow-up of ZTF-discovered TDEs. These data were taken in Full Frame mode with the thin filter using EPIC \citep{Struder2001} and are presented here for the first time. The observation data files were reduced using the \xmm Standard Analysis Software \citep[SAS;][]{Gabriel_04}. The raw data were processed with \texttt{epproc}, and only pn data were analyzed due to its superior sensitivity. Background flares were identified by inspecting 10--12\,keV light curves and used to define good-time intervals. Events were filtered to include only single and double patterns (\texttt{PATTERN<=4}). Source and background spectra were extracted from circular regions of radii $r_{\rm src}=35^{\prime\prime}$ and $r_{\rm bkg}=108^{\prime\prime}$ on the same CCD. The ARFs and RMFs were generated using \texttt{arfgen} and \texttt{rmfgen}, respectively.

The \xmm spectra were binned such that each bin has a background subtracted $\rm S/N= 10$ (\texttt{grouptype=snmin, groupscale=10} in \texttt{ftgrouppha}), given that the spectra have respectively, from E1 to E3, $\sim$ 36,000, 5,000, and 5,000 counts, this binning guarantees the spectra are in a Gaussian regime. 

\section{Additional Figures}\label{app:post}

\begin{figure*}[]
    \centering
    \includegraphics[width=0.95\linewidth]{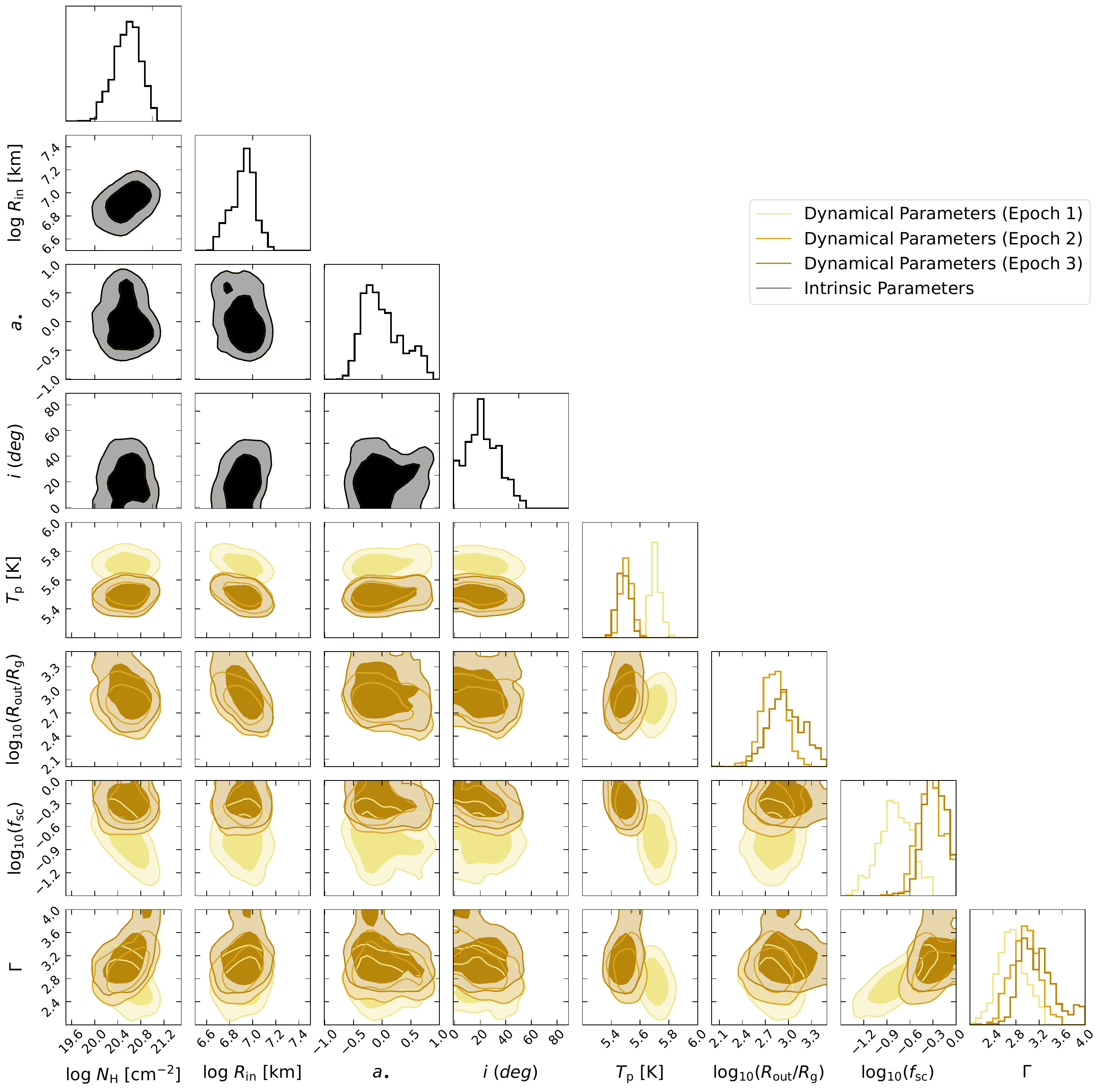}
    \caption{Full parameter posterior for the free parameters  our three epoch SED fitting. Intrinsic parameters are shown in black, while dynamical parameters are shown in colors. Contour represent 68\% and 95\% credible interval of the the projected two parameters.}
    \label{fig:5}
\end{figure*}

\begin{figure}[]
    \centering
    \includegraphics[width=0.95\linewidth]{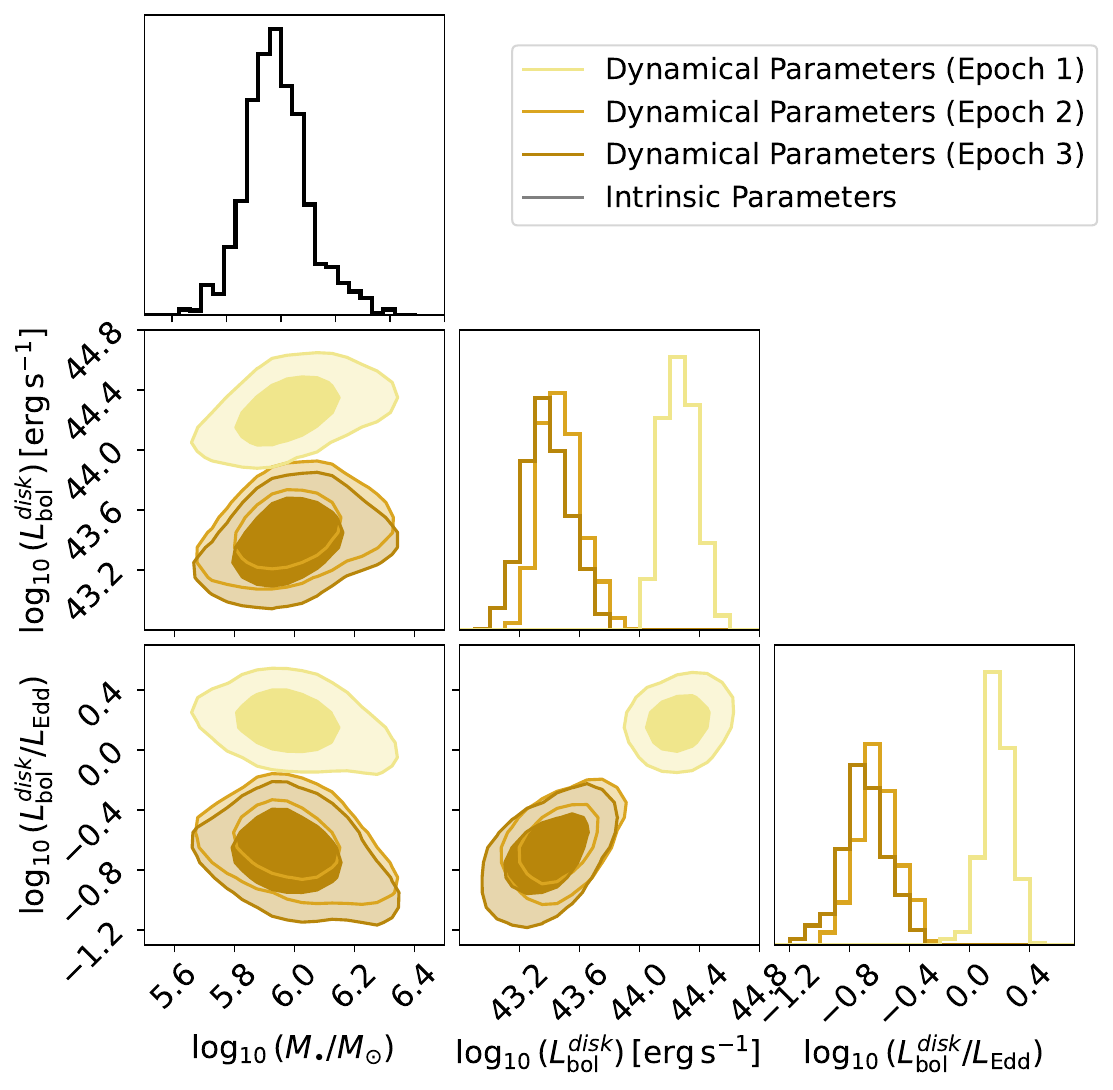}
    \caption{Full parameter posterior for secondary/derived parameters. Contours and labels as in Fig.~\ref{fig:5}.}
    \label{fig:6}
\end{figure}

In Fig.~\ref{fig:5} we show the full parameter posterior for the free parameters in our SED fitting, while in  Fig.~\ref{fig:6} we show the posterior on secondary derived parameters.

\bibliography{tde}{}
\bibliographystyle{aasjournal}

\end{document}